\documentclass[prd,aps,showpacs,nofootinbib]{revtex4}%
\usepackage{amsmath}
\usepackage{amsfonts}
\usepackage{amssymb}
\usepackage{bm}
\usepackage{graphicx}%
\providecommand{\U}[1]{\protect\rule{.1in}{.1in}}

\begin{document}
\title{Consistency analysis of a nonbirefringent Lorentz-violating planar model }
\author{Rodolfo Casana, Manoel M. Ferreira Jr, Roemir P. M. Moreira}
\affiliation{Departamento de F\'{\i}sica, Universidade Federal do Maranh\~{a}o (UFMA),
Campus Universit\'{a}rio do Bacanga, S\~{a}o Lu\'{\i}s - MA, 65085-580, Brazil.}

\begin{abstract}
In this work analyze the physical consistency of a nonbirefringent
Lorentz-violating planar model via the analysis of the pole structure of its
Feynman's propagators. The nonbirefringent planar model, obtained from the
dimensional reduction of the CPT-even gauge sector of the standard model
extension, is composed of a gauge and a scalar fields, being affected by
Lorentz-violating (LIV) coefficients encoded in the symmetric tensor
$\kappa_{\mu\nu}$. The propagator of the gauge field is explicitly\textbf{\ }%
evaluated and expressed in terms of linear independent symmetric tensors,
presenting only one physical mode. The same holds for the scalar propagator. A
consistency analysis is performed based on the poles of the propagators. The
isotropic parity-even sector is stable, causal and unitary mode for
$0\leq\kappa_{00}<1$. On the other hand, the anisotropic sector is stable and
unitary but in general noncausal. Finally, it is shown that this planar model
interacting with a $\lambda|\varphi|^{4}-$Higgs field supports compactlike
vortex configurations.

\end{abstract}

\maketitle

\section{Introduction}

The gauge sector of the of the Standard Model Extension (SME)
\cite{Samuel,Colladay} embraces a CPT-odd \cite{Jackiw} and a CPT-even
\cite{KM1,KM2,KM3,Kostelec} subsectors, which have been examined in several
aspects concerning supersymmetry \cite{Susy}, vacuum Cherenkov radiation
emission \cite{Cerenkov1,Cherenkov2}, radiative corrections \cite{Radiative},
Casimir effect \cite{Casimir}, anisotropies of the Cosmic Microwave Background
Radiation \cite{CMBR}, between other interesting issues \cite{Petrov}. Some
work was devoted to examine consistency aspects of the CPT-odd subsector
\cite{Adam,Soldati} and of the CPT-even one \cite{Prop2,Klinkmicro}.

The CPT-even gauge sector of the SME is represented by the CPT-even tensor
Lagrangian%
\begin{equation}
\mathit{{\mathcal{L}}}_{(1+3)}=-\frac{1}{4}F_{\hat{\mu}\hat{\nu}}F^{\hat{\mu
}\hat{\nu}}-\frac{1}{4}\left(  K_{F}\right)  _{\hat{\mu}\hat{\nu}\hat{\lambda
}\hat{\kappa}}F^{\hat{\mu}\hat{\nu}}F^{\hat{\lambda}\hat{\kappa} }, \label{L1}%
\end{equation}
with $\left(  K_{F}\right)  _{\alpha\beta\mu\nu}$ being the Lorentz-violating
tensor composed of 19 coefficients, with nine nonbirefringent and ten
birefringent ones, endowed with the symmetries of the Riemann tensor, and a
double null trace, $\left(  K_{F}\right)  ^{\hat{\mu}\hat{\nu}}{}_{\hat{\mu
}\hat{\nu}}=0.$\ The effects of this CPT-even electrodynamics on
fermion-fermion interaction\ was considered in
Refs.\cite{Klink2,Klink3,Interact1,Interact2,Interact3}.

A Lorentz-violating and\ CPT-even model in (1+2)-dimensions
\cite{Reduction2011} was first derived by means of the dimensional reduction
of the CPT-even Lagrangian (\ref{L1}), leading to the following planar
Lagrangian
\begin{equation}
\mathit{{\mathcal{L}}}_{(1+2)}=-\frac{1}{4}F_{\mu\nu}F^{\mu\nu}-\frac{1}%
{4}Z_{\mu\nu\lambda\kappa}F^{\mu\nu}F^{\lambda\kappa}+\frac{1}{2}\partial
_{\mu}\phi\partial^{\mu}\phi-C_{\mu\lambda}{\partial^{\mu}\phi\partial
^{\lambda}\phi}+T_{\mu\lambda\kappa}{\partial^{\mu}\phi}F^{\lambda\kappa
},\label{LP1}%
\end{equation}
where $Z_{\mu\nu\lambda\kappa},C_{\mu\lambda},T_{\mu\lambda\kappa}$ are
Lorentz-violating tensors which present together 19 components and affect the
electromagnetic and scalar sectors. Further investigations concerning Lorentz
violation in (1+2) dimension focused on the nonbirefringent CPT-even
electrodynamics of the SME, governed by the Lagrangian
\begin{equation}
\mathit{{\mathcal{L}}}_{(1+3)}=-\frac{1}{4}F_{\hat{\mu}\hat{\nu}}F^{\hat{\mu
}\hat{\nu}}-\frac{1}{2}\kappa_{\hat{\nu}\hat{\rho}}F_{\hat{\lambda}}{}%
^{\hat{\nu}}F^{\hat{\lambda}\hat{\rho}},\label{L2}%
\end{equation}
whose components are parametrized in the symmetric and traceless tensor,
$\kappa_{\hat{\nu}\hat{\rho}}.$ Its dimensional reduction provided the
following planar Lagrangian \cite{ReductionB2011}:
\begin{equation}
\mathcal{L}_{1+2}=-\frac{1}{4}F_{\mu\nu}F^{\mu\nu}-\frac{1}{2}\kappa_{\nu\rho
}F_{\lambda}^{\text{ \ }\nu}F^{\lambda\rho}+\frac{1}{2}[1-\eta]\partial_{\mu
}\phi\partial^{\mu}\phi+\frac{1}{2}\kappa_{\nu\rho}\partial^{\nu}\phi
\partial^{\rho}\phi+S_{\nu}F^{\nu\lambda}\partial_{\lambda}\phi,\label{LP2}%
\end{equation}
influenced by 9 Lorentz-violating coefficients contained in the tensors
$\kappa_{\nu\rho},$ $S_{\nu}.$ This planar model is composed of parity-odd and
parity-even components, allowing potential applications in general planar
systems \cite{Planar}. Both planar Lagrangian densities (\ref{LP1} and
\ref{LP2}) present scalar sectors endowed with a noncanonical kinetic term,
which has been used to study topological defects in (1+1) dimensions
\cite{Defects}, acoustic black holes with Lorentz-violation in (1+2)
dimensions \cite{Brito} and also the Bose-Einstein condensation of a bosonic
ideal gas \cite{Casana}.

In this work, we finalize the investigation concerning consistency aspects of
the planar model described by Lagrangian (\ref{LP2}), whose properties were
first analyzed in Ref. \cite{ReductionB2011}. With this purpose, we exactly
evaluate the Feynman's propagators for the electromagnetic and scalar sectors.
Such calculation, for the gauge field, is performed by means of a closed set
of tensors (non necessarily linearly independents), previously defined in Ref.
\cite{Prop2}. After expressing the propagator in terms of linearly independent
tensors, we proceed writing the dispersion relations for this planar theory
and comparing them with the relations of Ref. \cite{ReductionB2011}. The
physical dispersion relations are used to examine the causality, stability and
unitarity. It is shown that this planar theory is endowed with causality
illness, being stable and unitary for some known range of parameters. As a
clear physical motivation to study this kind of planar model, we show that it
constitutes a suitable framework for studying nonconventional vortex
solutions. Indeed, it allows the existence of compactlike uncharged BPS vortex
configurations, whenever coupled with the suitable a $\lambda|\varphi|^{4}%
-$Higgs potential.

This work is outlined as follows. In Sec. II, we explicitly carry out the
Feynman's propagators using a suitable parametrization for the LIV tensor. In
Sec. III, we analyze by using the dispersion relations, the stability,
causality and unitarity issues concerning this theory. In Sec. IV, we discuss
the {existence} of BPS vortex configurations. In Sec. V, we present our Conclusions.

\section{Feynman propagator for the planar theory}

It we disregard the mixing term ($S_{\mu}=0)$ in the Lagrangian (\ref{LP2}),
the gauge and scalar sectors turns out uncoupled, being governed by
\begin{equation}
\mathcal{L}_{1+2}=-\frac{1}{4}F_{\mu\nu}F^{\mu\nu}-\frac{1}{2}\kappa_{\nu\rho
}F_{\lambda}^{\text{ \ }\nu}F^{\lambda\rho}-\frac{1}{2\xi}(\partial_{\mu
}A^{\mu})^{2}+\frac{1}{2}[1-\eta]\partial_{\mu}\phi\partial^{\mu}\phi+\frac
{1}{2}\kappa_{\nu\rho}\partial^{\nu}\phi\partial^{\rho}\phi, \label{LP3}%
\end{equation}
where we have introduced the gauge-fixing term $(\partial_{\mu}A^{\mu}%
)^{2}/2\xi$, and the parameter $\eta$ given by
\begin{equation}
\eta=\kappa_{00}-\text{tr}\left(  \kappa_{ij}\right)  . \label{eta1}%
\end{equation}
Such Lagrangian can be written as
\begin{equation}
\mathcal{L}_{1+2}=\frac{1}{2}A_{\mu}\left[  D^{\mu\nu}\right]  A_{\nu}%
+\frac{1}{2}\phi\left[  \boxdot\right]  \phi,
\end{equation}
where the operator $D^{\mu\nu}$ defines the kinetic term of the gauge field
\begin{equation}
D^{\mu\nu}=\left(  \square+\kappa^{\alpha\beta}\partial_{\alpha}%
\partial_{\beta}\right)  g^{\mu\nu}+\lambda\partial^{\mu}\partial^{\nu
}+\square\kappa^{\mu\nu}-\kappa^{\mu\alpha}\partial_{\alpha}\partial^{\nu
}-\kappa^{\nu\alpha}\partial_{\alpha}\partial^{\mu}, \label{dd}%
\end{equation}
where $\lambda=\xi^{-1}-1$, and $\boxdot$ the scalar kinetic term
\begin{equation}
\boxdot=-(1-\eta)\square-\kappa_{\rho\sigma}\partial^{\rho}\partial^{\sigma}.
\label{KSc}%
\end{equation}
In this situation, we initiate evaluating the Feynman propagator for the gauge
sector, which fulfills the relation
\begin{equation}
D^{\mu\beta}\Delta_{\beta\nu}\left(  x-y\right)  =\delta{}^{\mu}{}_{\nu}%
\delta\left(  x-y\right)  . \label{Inv}%
\end{equation}

The gauge propagator is defined as $i\Delta_{\mu\nu}\left(  x-y\right)
=\left\langle 0\left\vert TA_{\mu}\left(  x\right)  A_{\nu}\left(  y\right)
\right\vert 0\right\rangle ,$ where $\Delta_{\mu\nu}\left(  x-y\right)  $ is
the Green's function of the operator $D^{\mu\nu}$ that appears in Eq.
(\ref{dd}). \ Working in the Fourier representation, we get
\begin{equation}
\widetilde{D}^{\mu\nu}(p)=-\left(  p^{2}+\kappa^{\alpha\beta}p_{\alpha
}p_{\beta}\right)  g^{\mu\nu}-\lambda p^{\mu}p^{\nu}-p^{2}\kappa^{\mu\nu
}+\kappa^{\mu\alpha}p_{\alpha}p^{\nu}+\kappa^{\nu\alpha}p_{\alpha}p^{\mu
}.\label{S2}%
\end{equation}
is a suitable way for inverting this tensor, which amounts at finding a tensor
$\tilde{\Delta}_{\mu\nu}$ that fulfills $\tilde{D}^{\mu\beta}\tilde{\Delta
}_{\beta\nu}=\delta{}^{\mu}{}_{\nu}$. Thus, we use the following
parametrization for a symmetric tensor,%
\begin{equation}
\kappa^{\mu\nu}=\frac{1}{2}(A^{\mu}B^{\nu}+A^{\nu}B^{\mu}),\label{Symm}%
\end{equation}
with nonnull trace, $\kappa^{\alpha}{}_{\alpha}=A\cdot B\neq0,$ and $A^{\mu
},B^{\nu}$ are two arbitrary three-vectors which comprise the
Lorentz-violating coefficients. Moreover, we have
\begin{equation}
\kappa^{00}=A^{0}B^{0},\ \kappa^{0i}=\frac{1}{2}(A^{0}B^{i}+A^{i}B^{0}),\text{
}\kappa^{ij}=\frac{1}{2}(A^{i}B^{j}+A^{j}B^{i}),
\end{equation}
here, $\kappa^{00}$ is the isotropic parity-even parameter\textbf{, }%
$\kappa^{0i}$ and $\kappa^{ij}$ are the anisotropic LIV coefficients retaining
the others parity-even and parity-odd ones. Replacing the parametrization
(\ref{Symm}) in Eq.(\ref{S2}), we have:
\begin{align}
\widetilde{D}^{\mu\nu} &  =-\left[  p^{2}+(p\cdot A)(p\cdot B)\right]
g^{\mu\nu}-\lambda p^{\mu}p^{\nu}-\frac{1}{2}p^{2}\left(  A^{\mu}B^{\nu
}+A^{\nu}B^{\mu}\right)  \nonumber\\[-0.3cm]
& \\
&  +\frac{1}{2}(p\cdot A)\left(  p^{\mu}B^{\nu}+p^{\nu}B^{\mu}\right)
+\frac{1}{2}(p\cdot B)\left(  p^{\mu}A^{\nu}+p^{\nu}A^{\mu}\right)  .\nonumber
\end{align}
In order to compute the inverse of $\widetilde{D}^{\mu\nu}$, we use the closed
algebra satisfied by the following tensors:
\begin{equation}
\Theta_{\mu\nu},\text{ }\omega_{\mu\nu},\text{ }A_{\mu}B_{\nu},~A_{\nu}B_{\mu
},\text{ }p_{\mu}A_{\nu},~p_{\nu}A_{\mu},\text{ }p_{\mu}B_{\nu},~p_{\nu}%
B_{\mu},~A_{\mu}A_{\nu},~B_{\mu}B_{\nu},\label{project}%
\end{equation}
which is presented in the Tables of Ref. \cite{Prop2}. Here, $\Theta_{\mu\nu
}=g_{\mu\nu}-\omega_{\mu\nu}$ and$\ \omega_{\mu\nu}=p_{\mu}p_{\nu}/p^{2}$ are
the transverse and longitudinal projectors, respectively. Due to the tensor
$\widetilde{D}^{\mu\nu}$ is symmetric in the Lorentz indices and under the
exchange $A\leftrightarrow B$, the gauge field propagator is expressed as
\begin{align}
\widetilde{\Delta}_{\mu\nu}\left(  p\right)   &  =\alpha_{1}\ \Theta_{\mu\nu
}+\alpha_{2}\ \omega_{\mu\nu}+\alpha_{3}A_{\mu}A_{\nu}+\alpha_{4}\left(
A_{\mu}B_{\nu}+A_{\nu}B_{\mu}\right)  \nonumber\\[-0.15cm]
& \label{PropF-0}\\[-0.15cm]
&  +\alpha_{5}\ \left(  p_{\mu}A_{\nu}+p_{\nu}A_{\mu}\right)  +\alpha
_{6}B_{\mu}B_{\nu}+\alpha_{7}\left(  p_{\mu}B_{\nu}+p_{\nu}B_{\mu}\right)
,\nonumber
\end{align}
with the coefficients $\alpha_{i}$ given as
\begin{align}
\alpha_{1} &  =-\frac{1}{p^{2}+\left(  A\cdot p\right)  \left(  B\cdot
p\right)  }~\ \ ,~\ \ ~\ \alpha_{3}=~\frac{\left(  B\cdot p\right)  ^{2}%
-p^{2}B^{2}}{4\boxtimes\left[  p^{2}+\left(  A\cdot p\right)  \left(  B\cdot
p\right)  \right]  }\\[0.3cm]
\alpha_{2} &  =\frac{2\left(  A\cdot B\right)  \left(  A\cdot p\right)
\left(  B\cdot p\right)  -A^{2}\left(  B\cdot p\right)  ^{2}-B^{2}\left(
A\cdot p\right)  ^{2}}{4\boxtimes\left[  p^{2}+\left(  A\cdot p\right)
\left(  B\cdot p\right)  \right]  }+\frac{\left(  A\cdot p\right)  \left(
B\cdot p\right)  }{p^{2}\boxtimes}-\frac{\xi}{p^{2}},\\[0.3cm]
\alpha_{4} &  =\frac{p^{2}\left[  2+\left(  A\cdot B\right)  \right]  +\left(
A\cdot p\right)  \left(  B\cdot p\right)  }{4\boxtimes\left[  p^{2}+\left(
A\cdot p\right)  \left(  B\cdot p\right)  \right]  }~\ \ ,~\ \ ~\ \alpha
_{6}=\frac{\left(  A\cdot p\right)  ^{2}-p^{2}A^{2}}{4\boxtimes\left[
p^{2}+\left(  A\cdot p\right)  \left(  B\cdot p\right)  \right]  },\\[0.3cm]
\alpha_{5} &  =-\frac{p^{2}\left[  2\left(  B\cdot p\right)  +\left(  A\cdot
B\right)  \left(  B\cdot p\right)  -B^{2}\left(  A\cdot p\right)  \right]
+2\left(  A\cdot p\right)  \left(  B\cdot p\right)  ^{2}}{4p^{2}%
\boxtimes\left[  p^{2}+\left(  A\cdot p\right)  \left(  B\cdot p\right)
\right]  },\ \\[0.3cm]
\alpha_{7} &  =-\frac{p^{2}\left[  2\left(  A\cdot p\right)  +\left(  A\cdot
B\right)  \left(  A\cdot p\right)  -A^{2}\left(  B\cdot p\right)  \right]
+2\left(  B\cdot p\right)  \left(  A\cdot p\right)  ^{2}}{4p^{2}%
\boxtimes\left[  p^{2}+\left(  A\cdot p\right)  \left(  B\cdot p\right)
\right]  },
\end{align}
where $A^{2}=A_{\mu}A^{\mu},$ $B^{2}=B_{\mu}B^{\mu}$, and the denominator
element $\boxtimes$ is
\begin{equation}
\boxtimes(p)=f(A,B)p^{2}+\left(  A\cdot p\right)  \left(  B\cdot p\right)
+\frac{1}{2}\left(  A\cdot B\right)  \left(  A\cdot p\right)  \left(  B\cdot
p\right)  +\frac{1}{4}B^{2}\left(  A\cdot p\right)  ^{2}+\frac{1}{4}%
A^{2}\left(  B\cdot p\right)  ^{2},\label{Caixa}%
\end{equation}%
\begin{equation}
f(A,B)=1+(A\cdot B)+\frac{1}{4}(A\cdot B)^{2}-\frac{1}{4}B^{2}A^{2}.
\end{equation}

At first sight, the gauge field propagator presents two simple poles in
$\boxtimes=0$\ and $p^{2}+\left(  A\cdot p\right)  \left(  B\cdot p\right)
=0,$\ and a double pole in $p^{2}=0$. However, as it will be shown below, the
gauge field presents only a unique physical pole, $\boxtimes=0,$\ that is
related to its physical dispersion relation. The second pole, $p^{2}+\left(
A\cdot p\right)  \left(  B\cdot p\right)  =0$, appears due the fact the gauge
propagator (\ref{PropF-0}) is expressed in terms of seven symmetric and
linearly dependent tensors arranged from the ten original tensors given in Eq.
(\ref{project}). It is well known that any second order symmetric tensor in
(2+1)-D can be written in terms of only six symmetric and linearly independent
(L.I.)\ ones. By choosing the set\ of L.I. tensors as being $\left\{
g^{\mu\nu},~\omega_{\mu\nu},~A_{\mu}A_{\nu},~A_{\mu}B_{\nu}+A_{\nu}B_{\mu
},~p_{\mu}A_{\nu}+p_{\nu}A_{\mu},~p_{\mu}B_{\nu}+p_{\nu}B_{\mu}\right\}  $, we
can express the tensor $B^{\mu}B^{\nu},$\ in terms of this L.I. set, in the
form%
\begin{equation}
B^{\mu}B^{\nu}=\beta_{1}g^{\mu\nu}+\beta_{2}p^{2}\omega^{\mu\nu}+\beta
_{3}A^{\mu}A^{\nu}+\beta_{4}\frac{A^{\mu}B^{\nu}+A^{\nu}B^{\mu}}{2}+\beta
_{5}\frac{A^{\mu}p^{\nu}+A^{\nu}p^{\mu}}{2}+\beta_{6}\frac{B^{\mu}p^{\nu
}+B^{\nu}p^{\mu}}{2},
\end{equation}
where
\begin{align}
\beta_{1} &  =-\frac{p^{2}\left[  A^{2}B^{2}-\left(  A\cdot B\right)
^{2}\right]  +2\left(  A\cdot p\right)  \left(  A\cdot B\right)  \left(
B\cdot p\right)  -A^{2}\left(  B\cdot p\right)  ^{2}-B^{2}\left(  A\cdot
p\right)  ^{2}}{\left(  A\cdot p\right)  ^{2}-A^{2}p^{2}},~\ \\[0.3cm]
\beta_{2} &  =-\frac{\ \left(  A\cdot B\right)  ^{2}-A^{2}B^{2}}{\left(
A\cdot p\right)  ^{2}-A^{2}p^{2}},~\ \beta_{3}=-\frac{\ \left(  B\cdot
p\right)  ^{2}-B^{2}p^{2}}{\left(  A\cdot p\right)  ^{2}-A^{2}p^{2}}%
,~\ \beta_{4}=-\frac{\ 2\left(  A\cdot\,B\right)  \,p^{2}-2\left(
p\cdot\,A\right)  \left(  p\cdot\,B\right)  }{\left(  A\cdot p\right)
^{2}-A^{2}p^{2}},~\ \\[0.3cm]
\beta_{5} &  =-\frac{\ 2\left(  p\cdot A\right)  B^{2}-2\left(  A\cdot
\,B\right)  \left(  p\cdot\,B\right)  }{\left(  A\cdot p\right)  ^{2}%
-A^{2}p^{2}},~\ \beta_{6}=-\frac{\ 2\left(  p\cdot B\right)  A^{2}-2\left(
A\cdot B\right)  \left(  p\cdot A\right)  }{\left(  A\cdot p\right)
^{2}-A^{2}p^{2}},
\end{align}

By replacing the expression for $B^{\mu}B^{\nu}$ in Eq. (\ref{PropF-0}), and
after some considerable algebraic effort, we obtain:%
\begin{align}
\widetilde{\Delta}_{\mu\nu}\left(  p\right)   &  =-\frac{1+\left(  A\cdot
B\right)  }{\boxtimes}\Theta_{\mu\nu}+\left[  \frac{\left(  A\cdot p\right)
\left(  B\cdot p\right)  }{p^{2}\boxtimes}-\frac{\xi}{p^{2}}\right]
\omega_{\mu\nu}+\frac{A_{\mu}B_{\nu}+A_{\nu}B_{\mu}}{2\boxtimes}\nonumber\\
& \\
&  -\left(  B\cdot p\right)  \frac{A_{\mu}p_{\nu}+A_{\nu}p_{\mu}}%
{2p^{2}\boxtimes}-\left(  A\cdot p\right)  \frac{B_{\mu}p_{\nu}+B_{\nu}p_{\mu
}}{2p^{2}\boxtimes}\text{.}\nonumber
\end{align}
In this form, $\widetilde{\Delta}_{\mu\nu}$ is expressed in terms only five
L.I. symmetric tensors, involving just eight of the original tensor set
(\ref{project}). The equation above now can also be expressed in terms of the
tensor $\kappa^{\mu\nu}$\ as
\begin{equation}
\widetilde{\Delta}_{\mu\nu}\left(  p\right)  =-\frac{\left(  1+\kappa^{\alpha
}{}_{\alpha}\right)  }{\boxtimes}\Theta_{\mu\nu}+\left[  \frac{\kappa
^{\rho\sigma}p_{\rho}p_{\sigma}}{p^{2}\boxtimes}-\frac{\xi~}{p^{2}}\right]
\omega_{\mu\nu}+\frac{\kappa_{\mu\nu}}{\boxtimes}-\frac{p_{\mu}\kappa
_{\nu\sigma}p^{\sigma}+p_{\nu}\kappa_{\mu\sigma}p^{\sigma}}{p^{2}\boxtimes
},\label{prop0}%
\end{equation}
and%
\begin{equation}
\boxtimes=\left[  1+\kappa^{\alpha}{}_{\alpha}+\frac{1}{2}\left(
\kappa^{\alpha}{}_{\alpha}\right)  ^{2}-\frac{1}{2}\left(  \kappa^{\alpha
\beta}\kappa_{\alpha\beta}\right)  \right]  p^{2}+\kappa^{\mu\nu}p_{\mu}%
p_{\nu}+\kappa^{\mu\alpha}\kappa_{\alpha}{}^{\nu}p_{\mu}p_{\nu}.
\end{equation}
It can be explicitly verified that the expression (\ref{prop0}) is the inverse
of the tensor $\widetilde{D}^{\mu\nu}\left(  p\right)  $ appearing in
(\ref{S2}),\textbf{\ }for\textbf{\ }any symmetric $\kappa^{\mu\nu}$. Thus, the
propagator in momentum space reads
\begin{equation}
\left\langle 0\left\vert TA_{\mu}\left(  p\right)  A_{\nu}\left(  -p\right)
\right\vert 0\right\rangle =-\frac{i}{\boxtimes}\left[  \left(  1+\kappa
^{\alpha}{}_{\alpha}\right)  \Theta_{\mu\nu}-\left(  \frac{\kappa^{\rho\sigma
}p_{\rho}p_{\sigma}}{p^{2}}-\xi\frac{\boxtimes}{p^{2}}\right)  \omega_{\mu\nu
}-\kappa_{\mu\nu}+\frac{p_{\mu}\kappa_{\nu\sigma}p^{\sigma}+p_{\nu}\kappa
_{\mu\sigma}p^{\sigma}}{p^{2}}\right]  .\label{prop1}%
\end{equation}
We can observe that the propagator presents a simple pole in $\boxtimes=0$ and
a double pole in $p^{2}=0$. As it will be shown, the pole $\boxtimes=0$ is the
one that provides physical dispersion relations, while the pole $p^{2}=0$ is unphysical.

One should evaluate the Feynman propagator for the scalar sector as well,
which in momentum space fulfills
\begin{equation}
\left\langle 0\left\vert T\phi\left(  p\right)  \phi\left(  -p\right)
\right\vert 0\right\rangle =i\tilde{\Delta}(p),
\end{equation}
where $\tilde{\Delta}(p)$ satisfies $\tilde{\boxdot}\tilde{\Delta}=1$, and
$\tilde{\boxdot}$ \ is the momentum representation of the scalar kinetic term
defined in (\ref{KSc}) $\tilde{\Delta}^{-1}=(1-\eta)p^{2}+\kappa^{\rho\sigma
}p_{\rho}p_{\sigma}$. Thus, the scalar propagator reads%
\begin{equation}
\left\langle 0\left\vert T\phi\left(  p\right)  \phi\left(  -p\right)
\right\vert 0\right\rangle =\frac{i}{(1-\eta)p^{2}+\kappa^{\rho\sigma}p_{\rho
}p_{\sigma}}.
\end{equation}
Note that it has a simple pole in $\tilde{\boxdot}=0$.

\section{Dispersion relations}

The dispersion relations for the gauge sector are read off from the physical
poles of the gauge propagator, that is, $\boxtimes=0$, or explicitly
\begin{equation}
\left[  1+\kappa^{\alpha}{}_{\alpha}+\frac{1}{2}\left(  \kappa^{\alpha}%
{}_{\alpha}\right)  ^{2}-\frac{1}{2}\left(  \kappa^{\alpha\beta}\kappa
_{\alpha\beta}\right)  \right]  p^{2}+\kappa^{\mu\nu}p_{\mu}p_{\nu}%
+\kappa^{\mu\alpha}\kappa_{\alpha}{}^{\nu}p_{\mu}p_{\nu}=0, \label{DRP2}%
\end{equation}
while the dispersion relation for the scalar sector is extracted from the
scalar propagator,%
\begin{equation}
(1-\eta)p^{2}+\kappa^{\rho\sigma}p_{\rho}p_{\sigma}=0. \label{DRP3B}%
\end{equation}
These dispersion relations are the same ones that were explicitly computed in
Ref. \cite{ReductionB2011} starting from the vacuum-vacuum amplitude
transition. With both relations in hands we can analyze the energy stability,
causality and unitarity of this theory.

\subsection{Stability}

We initiate\ analyzing the isotropic parity-even sector characterized by the
parameter $\kappa_{00}$. The expression (\ref{DRP2}) yields only a unique
dispersion relation for the gauge field,%
\begin{equation}
p_{0}=\pm\frac{|\mathbf{p|}}{\sqrt{1+\kappa_{00}}}, \label{DR1A}%
\end{equation}
which stands for positive-energy modes being compatible with energy stability
for $\kappa_{00}>-1$. For the same nonnull LIV coefficient, the scalar
dispersion relation Eq. (\ref{DRP3B}) provides%
\begin{equation}
p_{0}=\pm|\mathbf{p|}\sqrt{1-\kappa_{00}}, \label{DR1S}%
\end{equation}
which stands for positive-energy modes being compatible with energy stability
for $\kappa_{00}<1$. Concerning the isotropic parameter, therefore, the entire
model will be stable for $\left\vert \kappa_{00}\right\vert <1$.

Taking now as nonnull only the anisotropic parameter $\kappa_{0i}$, the
dispersion relation (\ref{DRP2}) for the gauge field\textbf{\ is}
\begin{equation}
p_{0}=\kappa_{0i}p_{i}\pm|\mathbf{p|}\sqrt{1+\left(  \kappa_{0i}\right)  ^{2}%
}, \label{DR2B}%
\end{equation}
while for the scalar degree of freedom, Eq. (\ref{DRP3B}) yields%
\begin{equation}
p_{0}=\kappa_{0i}p_{i}\pm|\mathbf{p|}\sqrt{1+\frac{\left(  \kappa_{0i}%
p_{i}\right)  }{\mathbf{p}^{2}}^{2}}. \label{DR2S}%
\end{equation}
These dispersion relations\ represent modes compatible with energy stability
for any value of the LIV parameter $\kappa_{0i}$.

For nonnull $\kappa_{ij}$, the dispersion relation (\ref{DRP2}) for the gauge
degree of freedom takes the form%
\begin{equation}
p_{0}=\pm N_{0}\left\vert \mathbf{p}\right\vert \sqrt{1-\kappa_{ij}p_{i}%
p_{j}/\mathbf{p}^{2}}, \label{DRanisot}%
\end{equation}
with $N_{0}=\sqrt{(1-\text{tr}\left(  \kappa_{ij}\right)  )/(1-\text{tr}%
\left(  \kappa_{ij}\right)  +\det\left(  \kappa_{ij}\right)  )}$. From Eq.
(\ref{DRP3B}), the scalar field supports the dispersion relation%
\begin{equation}
p_{0}=\pm\left\vert \mathbf{p}\right\vert \sqrt{1+\left[  1+\text{tr}\left(
\kappa_{ij}\right)  \right]  ^{-2}\kappa_{ij}p_{i}p_{j}/\mathbf{p}^{2}}.
\label{DR3S}%
\end{equation}
\ These dispersion relations give modes supporting energy stability for
sufficiently small values of the parameter $\kappa_{ij}$.

\subsection{Causality analysis}

The poles of the Feynman propagator can be properly used to analyze the
causality of the associated theory \cite{Sexl}. In order to have a causal mode
(preventing the existence of tachyons), one must have $p^{2}\geq0$. A mode
confident criterium for defining causality is the evaluation of the group and
front velocities $(u_{g}=dp_{0}/d|\mathbf{p|},$ $u_{front}=\lim
_{|\mathbf{p|\rightarrow\infty}}u_{phase})$, so that causality is ensured if
$u_{g},u_{front}\leq1.$ In Refs.\cite{Prop2}, the causality of the original
four-dimensional CPT-even model was examined in its parity-odd and parity-even
subsectors, revealing the existence of noncausal modes.

As for the isotropic sector, the dispersion relations (\ref{DR1A}) and
(\ref{DR1S}) yield the group velocities%
\begin{equation}
u_{g}=\left(  1+\kappa_{00}\right)  ^{-1/2},~u_{g}=\left(  1-\kappa
_{00}\right)  ^{1/2},
\end{equation}
which lead to causal modes once the condition $0\leq\kappa_{00}\leq1$ is
obeyed (criterium for having $u_{g},u_{front}\leq1)$, otherwise the causality
is violated. \ As for the anisotropic modes ruled by $\kappa_{0i}$, we
consider the dispersion relations (\ref{DR2B}) and (\ref{DR2S}), whose
respective group velocities \ are
\begin{equation}
u_{g}=\sqrt{1+\left(  \kappa_{0i}\right)  ^{2}}\pm\left\vert \kappa
_{0i}\right\vert \cos\theta_{i},~u_{g}=\sqrt{1+\left\vert \kappa
_{0i}\right\vert ^{2}\cos^{2}\theta}\pm\left\vert \kappa_{0i}\right\vert
\cos\theta_{i}. \label{vg1}%
\end{equation}
where $\cos\theta_{i}=p_{i}/\left\vert \mathbf{p}\right\vert $. The
anisotropic modes associated with $\kappa_{ij}$ are described by the
dispersion relations (\ref{DRanisot}) and (\ref{DR3S}) which provide the
following group velocities%
\begin{equation}
u_{g}=N_{0}\sqrt{1-\kappa_{ij}p_{i}p_{j}/\mathbf{p}^{2}},~\ u_{g}%
=\sqrt{1+\left[  1+\text{tr}\left(  \kappa_{ij}\right)  \right]  ^{-2}%
\kappa_{ij}p_{i}p_{j}/\mathbf{p}^{2}}. \label{vg2}%
\end{equation}
\textbf{A}s the group velocities (\ref{vg1}-\ref{vg2}) may be larger than $1$
for some specific configurations of $\kappa_{0i},\kappa_{ij}$, we conclude
that these anisotropic modes are\ in general noncausal.

\subsection{Unitarity analysis}

The unitarity analysis of this model at tree-level is here carried out through
the saturation of the propagators with external currents \cite{Veltman}, which
must be implemented by means of the saturated propagator ($SP$), a scalar
quantity given as follows:
\begin{equation}
SP=J^{\ast\mu}\text{Res}(i\Delta_{\mu\nu})\text{ }J^{\nu},
\end{equation}
where Res$(i\Delta_{\mu\nu})$ is the matrix residue evaluated at the pole of
the propagator. The gauge current $(J^{\mu})$\ satisfies the conservation law
$\left(  \partial_{\mu}J^{\mu}=0\right)  ,$\ which in momentum space is read
as $p_{\mu}J^{\mu}=0$. In accordance with this method, the unitarity analysis
is assured whenever the imaginary part of the saturation $SP$\ (at the poles
of the propagator) is positive-definite. A way to carry out the saturation
consists of determining the residue of the propagator matrix, evaluated at its
own poles. Such analysis can state that a mode is physical or nonphysical
(whenever the saturation is positive or nonnegative), leading to a result
which should be reconciliated with the existence of only a physical mode.

From the propagator given in (\ref{prop1})\ the residue in the pole
$\boxtimes(p)=0$ is
\begin{equation}
\text{Res}\left(  i\tilde{\Delta}_{\mu\nu}\right)  =-\frac{g_{\mu\nu}\left(
1+\kappa^{\alpha}{}_{\alpha}\right)  }{\varrho}+\left[  \frac{\varrho
~\kappa^{\rho\sigma}p_{\rho}p_{\sigma}}{\left(  \Sigma-\mathbf{p}^{2}%
\varrho\right)  ^{2}}-\frac{\left(  1+\kappa^{\alpha}{}_{\alpha}\right)
}{\Sigma-\mathbf{p}^{2}\varrho}\right]  p_{\mu}p_{\nu}+\frac{\kappa_{\mu\nu}%
}{\varrho}-\frac{p_{\mu}\kappa_{\nu\sigma}p^{\sigma}+p_{\nu}\kappa_{\mu\sigma
}p^{\sigma}}{\Sigma-\mathbf{p}^{2}\varrho}, \label{Res1a}%
\end{equation}
where%
\begin{align}
\Sigma &  =2p_{0}\left[  \left(  1+\kappa_{00}\right)  \left(  \kappa
_{0i}p_{i}\right)  -\left(  \kappa_{0i}\kappa_{ij}p_{j}\right)  \right]
-\left[  1-\text{tr}\left(  \kappa_{ij}\right)  \right]  \left[  \left(
\kappa_{ij}p_{i}p_{j}\right)  -\mathbf{p}^{2}\left(  1+\kappa_{00}\right)
\right]  +\left(  \epsilon_{ij}\kappa_{0i}p_{j}\right)  ^{2},\nonumber\\
& \\
\varrho &  =\left(  1+\kappa_{00}\right)  \left[  1+\kappa_{00}-\text{tr}%
\left(  \kappa_{ij}\right)  \right]  +\det\left(  \kappa_{ij}\right)
.\nonumber
\end{align}
Then the saturated residue of the gauge propagator (taking into account the
current conservation) is%
\begin{equation}
SP=i\frac{J^{\mu}\kappa_{\mu\nu}J^{\nu}-J^{2}\left(  1+\kappa^{\alpha}%
{}_{\alpha}\right)  }{\varrho} \label{SAT1}%
\end{equation}

We will investigate the unitarity for three situations: (i) the isotropic
coefficient, $\kappa_{00},$ (ii) the anisotropic coefficients, $\kappa_{0i},$
(iii) the anisotropic coefficients, $\kappa_{ij}.$

We consider first the isotropic configuration, represented by $\kappa_{00}$.
In this case, $\varrho=(1+\kappa_{00})^{2}$\ and the saturation is
\begin{equation}
SP=i\frac{\left(  \mathbf{p\times J}\right)  ^{2}}{(1+\kappa_{00}%
)\mathbf{p}^{2}},
\end{equation}
which will imply a positive-definite imaginary part whenever $\kappa_{00}>-1
$. The isotropic parity-even sector gives an stable and unitary for
$|\kappa_{00}|<1,$

We consider now the anisotropic configuration associated with the coefficient
$\kappa_{0i}$. In this case, $\varrho=1$ and the saturation is%
\begin{equation}
SP=i\left[  \mathbf{J}^{2}-\left(  J_{0}\right)  ^{2}-2\kappa_{0i}J_{0}%
J_{i}\right]  .
\end{equation}
The saturation can be written as
\begin{equation}
\mathbf{J}^{2}-\left(  J_{0}\right)  ^{2}-2\kappa_{0i}J_{0}J_{i}%
=\frac{\mathcal{J}}{\left(  \mathbf{\Lambda}\cdot\mathbf{p}\pm\left\vert
\mathbf{p}\right\vert \sqrt{1+\mathbf{\Lambda}^{2}}\right)  ^{2}},
\end{equation}
where we have defined the vector $\Lambda^{i}=\kappa^{0i}\ $and
\begin{equation}
\mathcal{J}=\left(  \mathbf{p\times J}\right)  ^{2}+\mathbf{J}^{2}[\Lambda
^{2}\mathbf{p}^{2}-\left(  \mathbf{\Lambda\cdot p}\right)  ^{2}\mathbf{]}%
+2p_{0}[\left(  \mathbf{\Lambda\cdot p}\right)  \mathbf{J}^{2}-\left(
\mathbf{p\cdot J}\right)  \left(  \mathbf{\Lambda\cdot J}\right)  ],
\end{equation}
which can be rewritten by choosing the following orthonormal basis composed by
the vectors $\mathbf{u},~\mathbf{v}$:
\begin{equation}
\mathbf{u}=\frac{\mathbf{p}}{\left\vert \mathbf{p}\right\vert },~\ \mathbf{v}%
=\frac{\mathbf{\Lambda-}\left(  \mathbf{\Lambda\cdot u}\right)  \mathbf{u}%
}{\left\vert \mathbf{\Lambda-}\left(  \mathbf{\Lambda\cdot u}\right)
\mathbf{u}\right\vert },~\ \ \ \mathbf{\Lambda=}\left(  \kappa_{01}%
,\kappa_{02}\right)  .
\end{equation}
In this basis, we achieve%
\begin{equation}
\mathcal{J}=\left[  J_{\mathbf{u}}\Lambda_{\mathbf{v}}\left\vert
\mathbf{p}\right\vert -p_{0}J_{\mathbf{v}}\right]  ^{2},
\end{equation}
and we have $p=\left\vert \mathbf{p}\right\vert u,~\ \Lambda=\Lambda
_{\mathbf{u}}u+\Lambda_{\mathbf{v}}v,~\ J=J_{\mathbf{u}}u+J_{\mathbf{v}}v$.
Thus, the saturation is read
\begin{equation}
SP=i\left[  \frac{J_{\mathbf{u}}\Lambda_{\mathbf{v}}\left\vert \mathbf{p}%
\right\vert -p_{0}J_{\mathbf{v}}}{\mathbf{\Lambda}\cdot\mathbf{p}\pm\left\vert
\mathbf{p}\right\vert \sqrt{1+\mathbf{\Lambda}^{2}}}\right]  ^{2},
\end{equation}
which is positive-definite, compatible with the unitarity requirements.

We finally consider the anisotropic configuration represented by the
coefficients $\kappa_{ij}$. In this case, $\varrho=1-$tr$\left(  \kappa
_{ij}\right)  +\det\left(  \kappa_{ij}\right)  $,\ and the saturation is%
\begin{equation}
SP=i\frac{\kappa_{ij}J_{i}J_{j}+\left[  1-\text{tr}\left(  \kappa_{ij}\right)
\right]  \left[  \mathbf{J}^{2}-\left(  J_{0}\right)  ^{2}\right]
}{1-\text{tr}\left(  \kappa_{ij}\right)  +\det\left(  \kappa_{ij}\right)  }.
\end{equation}
The numerator can be written as%
\begin{equation}
\kappa_{ij}J_{i}J_{j}+\left[  1-\text{tr}\left(  \kappa_{ij}\right)  \right]
\left[  \mathbf{J}^{2}-\left(  J_{0}\right)  ^{2}\right]  =\frac{\left[
\epsilon_{ij}p_{i}J_{j}+\epsilon_{kj}\kappa_{ji}J_{k}p_{i}\right]  ^{2}%
}{\left[  \mathbf{p}^{2}-\left(  \kappa_{ij}p_{i}p_{j}\right)  \right]  },
\end{equation}
so that the saturation finally read
\begin{equation}
SP=i\frac{\left[  \epsilon_{ij}p_{i}J_{j}+\epsilon_{kj}\kappa_{ji}J_{k}%
p_{i}\right]  ^{2}}{\left[  \mathbf{p}^{2}-\kappa_{ij}p_{i}p_{j}\right]
\left[  1-\text{tr}\left(  \kappa_{ij}\right)  +\det\left(  \kappa
_{ij}\right)  \right]  },
\end{equation}
which is positive-definite for sufficiently small values of the background
$\kappa_{ij}$.

The residue in the double pole $p^{2}=0$ is proportional to $p_{\mu}p_{\nu}$,
and due to current conservation $\left(  p_{\mu}J^{\mu}=0\right)  ,$ its
saturation is zero.\ Thus, this is a nonphysical pole. Therefore, the unique
physical pole is $\boxtimes(p)=0$, which yields unitary modes, as examined above.

\section{One application: BPS vortex configurations}

In a recent work \cite{Carlisson}, it was demonstrated the existence of BPS
uncharged vortex configurations in the context of the nonbirefringent and
Lorentz-violating model of Ref. \cite{Prop2}, supplemented by a Higgs sector.
It {was} argued that the Lorentz-violating parameter modifies the profiles of
the {BPS vortex solutions} yielding a compactlike behavior similar to the ones
stemming from {the} k-field theories \cite{Compact}, \cite{kfield}. The
existence of vortices in the LIV framework of Ref. \cite{Carlisson} implies
{an expectation} for the observation of vortices in the context of the planar
model developed {here} when properly coupled to the Higgs field. In the
following {we show} that this is really the case.

The starting point in the investigation about vortex configurations is the
planar Lagrangian (\ref{LP3}), supplemented with the Higgs field, $\varphi,$
\begin{equation}
\mathcal{L}_{1+2}=-\frac{1}{4}F_{\mu\nu}F^{\mu\nu}-\frac{1}{2}\kappa_{\nu\rho
}F_{\lambda}^{\text{ \ }\nu}F^{\lambda\rho}+\frac{1}{2}[1-\eta]\partial_{\mu
}\phi\partial^{\mu}\phi+\frac{1}{2}\kappa_{\nu\rho}\partial^{\nu}\phi
\partial^{\rho}\phi+\left\vert \mathcal{D}_{\mu}\varphi\right\vert
^{2}-U\left(  \left\vert \varphi\right\vert ,\phi\right)  , \label{eqq1}%
\end{equation}
coupled to the gauge sector by the covariant derivative, $\left(
\mathcal{D}_{\mu}\varphi\right)  =\partial_{\mu}\varphi-ieA_{\mu}\varphi$, and
endowed with a suitable potential, $U\left(  \left\vert \varphi\right\vert
,\phi\right)  $, able to satisfy BPS conditions. The neutral scalar field
$\phi$, stemming from the dimensional reduction, plays a role analogue to the
auxiliary field $N$ inserted in the Maxwell-Chern-Simons-Higgs model
\cite{MCSV,Bolog} in order to provide BPS vortex solutions.

In the analysis of Ref. \cite{Carlisson}, we set $\kappa_{0i}=0$, which
restrains our study only to uncharged vortex solutions. In such circumstance,
one of the BPS equations of the full model states that%
\begin{equation}
A_{0}=\mp\phi=0,\label{bps_01}%
\end{equation}
The condition $A_{0}=0$\ is compatible with uncharged solutions, and satisfies
the Gauss's law\ trivially. Under this BPS condition, the potential becomes
\begin{equation}
U\left(  \left\vert \varphi\right\vert \right)  =\frac{1}{2\left(  1-s\right)
}\left(  ev^{2}-e\left\vert \varphi\right\vert ^{2}\right)  ^{2},
\end{equation}
where $v$ represents the vacuum expectation value of the Higgs field and
$s=\kappa_{ii}=\kappa_{11}+\kappa_{22}$.

This way, for uncharged solutions, the only nonnull stationary equations of
motion of the model (\ref{eqq1}) are the Ampere law
\begin{equation}
\left(  \epsilon_{aj}-\epsilon_{bj}\kappa_{ab}-\epsilon_{ab}\kappa
_{jb}\right)  \partial_{a}B{}-ie\left(  \varphi^{\ast}\partial_{j}%
\varphi-\varphi\partial_{j}\varphi^{\ast}\right)  -2e^{2}A_{j}\left\vert
\varphi\right\vert ^{2}=0,\label{Ampere1}%
\end{equation}
and the equations of motion for the Higgs field,
\begin{equation}
\nabla^{2}\varphi-2ieA_{j}\partial_{j}\varphi+e^{2}A_{0}^{2}\varphi-e^{2}%
A_{j}^{2}\varphi-e^{2}\phi^{2}\varphi-\frac{\partial U\left(  \left\vert
\varphi\right\vert \right)  }{\partial\varphi^{\ast}}=0.\label{Higgs1}%
\end{equation}

In this stationary regime, the canonical energy of the\ uncharged system is
written as%
\begin{equation}
E=\pm ev^{2}\int d^{2}xB+\int d^{2}x\left\{  \frac{1}{2}\left(  1-s\right)
\left[  B\mp\frac{ev^{2}-e\left\vert \varphi\right\vert ^{2}}{\left(
1-s\right)  }\right]  ^{2}+\left\vert D_{\pm}\varphi\right\vert ^{2}\mp
\frac{1}{2}\epsilon_{ij}\partial_{i}J_{j}\right\}  .
\end{equation}
Here, we have defined the operators
\begin{equation}
D_{\pm}\varphi=D_{1}\varphi\pm iD_{2}\varphi,
\end{equation}
and the current density $J_{j}=i\left(  \varphi^{\ast}\partial_{j}%
\varphi-\varphi\partial_{j}\varphi^{\ast}\right)  +2eA_{j}\left\vert
\varphi\right\vert ^{2}$.

In order to minimize the energy, we impose other two BPS\ conditions
\begin{align}
&  \displaystyle{D_{\pm}\varphi=0},\label{bps_02}\\[0.3cm]
&  \displaystyle{B=\pm\frac{ev^{2}-e\left\vert \varphi\right\vert ^{2}%
}{\left(  1-s\right)  }}. \label{bps_03}%
\end{align}
The three BPS conditions, (\ref{bps_01}), (\ref{bps_02}) and (\ref{bps_03}),
reduce the energy to be proportional to the magnetic flux,%
\begin{equation}
E_{BPS}=\pm ev^{2}\int d^{2}x~B,
\end{equation}
once the term $\epsilon_{ij}\partial_{i}J_{j}$ does not contribute to the
energy. It is worthwhile to note that the BPS equations reproduce the Ampere's
law (\ref{Ampere1}) and the stationary equation for the Higgs field
(\ref{Higgs1}). The vortex solutions are provided by the BPS equations,
(\ref{bps_02}) and (\ref{bps_03}), under appropriated boundary conditions.

In order to search for uncharged stable vortex configurations, we state the
usual ansatz for static rotationally symmetric vortex solutions, working in
polar coordinates $(r,\theta)$, and wherein the fields are parametrized as
\begin{equation}
A_{\theta}=-\frac{1}{er}\left[  a\left(  r\right)  -n\right]  ,~\ \ \varphi
=vg\left(  r\right)  e^{in\theta}, \label{Ans2}%
\end{equation}
where $a\left(  r\right)  ,$ $g\left(  r\right)  $ are regular scalar
functions satisfying appropriated boundary conditions that provide topological
vortex solutions possessing a finite total energy:
\begin{align}
g\left(  r\rightarrow\infty\right)   &  \rightarrow1\text{, \ \ }a\left(
r\rightarrow\infty\right)  \rightarrow0^{\pm},\label{bc1}\\
g\left(  r\rightarrow0\right)   &  \rightarrow0\;,\;\;a\left(  r\rightarrow
0\right)  \rightarrow n, \label{bc2}%
\end{align}
where the signal $+$\ corresponds to $n>0$\ and $-$\ to $n<0$,\ with
$n$\ being the winding number of the topological solution. In this ansatz the
magnetic field is
\begin{equation}
B\left(  r\right)  =-\frac{a^{\prime}}{er}, \label{Brot}%
\end{equation}
while the BPS equations, \ (\ref{bps_02}) and (\ref{bps_03}), are written as
\begin{align}
g^{\prime}  &  =\pm\frac{ag}{r},\label{BPS1}\\[0.2cm]
\frac{a^{\prime}}{er}  &  =\mp\frac{ev^{2}}{\left(  1-s\right)  }\left(
1-g^{2}\right)  . \label{BPS2}%
\end{align}

We note that these are the same BPS equations attained in Ref.
\cite{Carlisson}, so that the same results can be achieved here by numerical
simulations. We have thus shown that the present planar model supports
uncharged BPS solutions with a compactlike profile. Although this first
investigation has not unveiled a new BPS behavior in relation to the uncharged
profiles of Ref. \cite{Carlisson}, it confirms that this planar model supports
vortex solutions which could find application in planar systems where the
spatial extension of the defect depends on some additional parameter.

Another interesting investigation is to analyze the existence of charged
vortices in the planar Lorentz-violating model of Lagrangian (\ref{eqq1}) (in
the absence of the Chern-Simons term), such possibility is supported by its
Gauss's law
\begin{equation}
\left(  1+\kappa_{00}\right)  \partial_{j}\partial_{j}A_{0}-\kappa
_{ij}\partial_{i}\partial_{j}A_{0}{}+\epsilon_{ij}\kappa_{0i}\partial_{j}%
B{}-2e^{2}A_{0}\left\vert \phi\right\vert ^{2}=0. \label{Gauss_1}%
\end{equation}
which explicitly shows that a nonnull $\kappa_{0i}$ LIV parameter mixes the
electric and magnetic sectors of the model, opening the possibility for
existence of charged solutions.

\section{Conclusions}

In this work, we have continued examining the nonbirefringent CPT-even planar
electrodynamics stemming from the dimensional reduction of the CPT-even gauge
sector of the SME, first derived in Ref. \cite{ReductionB2011}. More
specifically, we have explicitly evaluated the Feynman propagator for this
planar model and analyzed some of its properties. The Feynman's propagator of
the electromagnetic sector was carried out using a set of ten tensors forming
a closed algebra.\ In the end, the gauge propagator was expressed in terms of
only eight of these tensors, arranged in such a way to yield five symmetric
linear independent tensors.\textbf{\ }The propagator poles yielded dispersion
relations which are compatible with the ones of Ref. \cite{ReductionB2011} and
revealed the existence of only one degree of freedom for each propagator, as
it is usual in a (1+2)-dimensional electrodynamics. The causality and
unitarity of electromagnetic and scalar sectors were examined. Such analysis
has shown that the isotropic parity-even sector is stable and unitary for
$|\kappa_{00}|<1,$ but causal for $0\leq\kappa_{00}\leq1$. \ So, this sector
is stable, causal and unitary for $0\leq\kappa_{00}<1$. On the other hand, the
anisotropic sector\textbf{\ }ruled by $\kappa_{0i}$ is stable and unitary for
arbitrary values of $\kappa_{0i}$, however it can be noncausal for the some
configurations of $\kappa_{0i}$. Finally, the anisotropic sector characterized
by $\kappa_{ij}$ is stable and unitary for sufficiently small values of the
LIV coefficients, but it is in general noncausal. This behavior is similar
with the scenario of the original (1+3)-dimensional theory \cite{Prop2} from
which this planar model was derived.

Finally, we have shown that the planar model examined in this work can support
uncharged BPS\ vortex solutions very similar to the ones of Ref.
\cite{Carlisson}, which can find applications in condensed matter system where
this Lorentz-violating electrodynamics may be interpreted as an effective
theory describing, for example, anisotropies or impurities of a particular
material. This first study also raises the issue about the existence of BPS
planar charged vortex solutions (in the absence of the Chern-Simons terms). As
one knows, the attainment of charged vortices is only possible when the
Chern-Simons term is considered \cite{CSV,Bolog}, once this term couples the
electric and magnetic sectors of the theory. In the present model, such
coupling is played by the Lorentz-violating parameter $\kappa_{0i}$, which
implies the inclusion of the term $\epsilon_{ij}\kappa_{0i}\partial_{j}B{}%
$\ in the Gauss's law of the full model (\ref{Gauss_1}). With it, we achieve a
richer model which certainly engenders charged planar BPS vortex
configurations. This investigation is under finalization \cite{Carlisson2} and
will be reported elsewhere.

\begin{acknowledgments}
The authors are grateful to FAPEMA, CAPES and CNPq (Brazilian research
agencies) for invaluable financial support.
\end{acknowledgments}

\end{document}